\documentclass[onecolumn, letterpaper, 10 pt, doublespace]{IEEEtran}  % Comment this line out if you need a4paper

\IEEEoverridecommandlockouts                              % This command is only needed if
\expandafter\def\expandafter\normalsize\expandafter{%
    \normalsize
    \setlength\abovedisplayskip{3pt}
    \setlength\belowdisplayskip{3pt}
    \setlength\abovedisplayshortskip{3pt}
    \setlength\belowdisplayshortskip{3pt}
}

\usepackage{graphicx}
\usepackage[perpage]{footmisc}
\usepackage{caption}
\usepackage{tabularx}
\usepackage{color}
\usepackage{wrapfig}
\usepackage{epsfig} % for postscript graphics files
\usepackage{mathptmx} % assumes new font selection scheme installed
\usepackage{times} % assumes new font selection scheme installed
\usepackage{amsmath} % assumes amsmath package installed
\usepackage{amssymb}  % assumes amsmath package installed
\usepackage{cite} % pachage for citations
\usepackage[space]{grffile}
\usepackage{epstopdf}
\usepackage{pdfpages}
\usepackage[mathscr]{euscript}
\usepackage[utf8]{inputenc}
\usepackage{lipsum}
\usepackage[normalem]{ulem} % either use this (simple) or
\usepackage{algorithm,algorithmic}
\DeclareMathOperator*{\argmin}{arg\,min}

\title{Sub-optimal Tracking in Switched Systems with Controlled Subsystems and Fixed-mode Sequence using Approximate Dynamic Programming }

\author{Tohid Sardarmehni$^{1}$ and Xingyong Song $^{2,*}$
    \thanks{$^{1}$ Tohid Sardarmehni, Postdoctoral Research Fellow,	Department of Engineering Technology and Industrial Distribution, Texas A\&M University, College Station, TX, 77843, USA.
{\tt tsardarmehni@smu.edu}}%

\thanks{$^{2,*}$ Xingyong Song, Assistant Professor, Department of Engineering Technology and Industrial Distribution; Department of Mechanical Engineering; College of Engineering; 	Texas A\&M University, College Station, TX, 77843, USA (Corresponding Author).
	  {\tt songxy@tamu.edu}
    }
\thanks{This research was partially supported by the National Academies of Sciences, Engineering and Medicine GRP Early Career Research Fellow Award. }
\thanks{This article has been accepted for oral presentation at 2019 Dynamic System and Control Conference. The content is the same as the final edition of the accepted paper. However, the presentation might be different. }
}
\usepackage{graphicx}
\usepackage{tabularx}
\usepackage{color}
\usepackage{wrapfig}
\usepackage{epsfig} % for postscript graphics files
\usepackage{mathptmx} % assumes new font selection scheme installed
\usepackage{times} % assumes new font selection scheme installed
\usepackage{amsmath} % assumes amsmath package installed
\usepackage{amssymb}  % assumes amsmath package installed
\usepackage{cite} % pachage for citations
\usepackage[space]{grffile}
\usepackage{epstopdf}
\usepackage[mathscr]{euscript}
\usepackage[utf8]{inputenc}
\usepackage{lipsum}
\usepackage{mathtools}
\usepackage[margin=0.5cm]{caption}
\usepackage{graphicx}
\usepackage{caption}
\usepackage{tabularx}
\usepackage{wrapfig}
\usepackage{epsfig} % for postscript graphics files
\usepackage{mathptmx} % assumes new font selection scheme installed
\usepackage{times} % assumes new font selection scheme installed
\usepackage{amsmath} % assumes amsmath package installed
\usepackage{amssymb}  % assumes amsmath package installed
\usepackage{cite} % pachage for citations
\usepackage[space]{grffile}
\usepackage{epstopdf}
\usepackage[utf8]{inputenc}
\usepackage{color}
\usepackage{slashbox}
\usepackage{lipsum}
\usepackage{float}
\usepackage{placeins}
\usepackage[normalem]{ulem} % either use this (simple) or
\usepackage{algorithm}
\usepackage{algorithmic}
\usepackage[algo2e]{algorithm2e} 
\usepackage{setspace}
\usepackage{mathrsfs}

\DeclareMathOperator{\diag}{diag}

\newtheorem{Rem}{Remark}
\newtheorem{Def}{Definition}

\newtheorem{Assumption}{Assumption}

\begin{document}
\maketitle    
%\thispagestyle{empty}
%\pagestyle{empty}

%%%%%%%%%%%%%%%%%%%%%%%%%%%%%%%%%%%%%%%%%%%%%%%%%%%%%%%%%%%%%%%%%%%%%%
\begin{abstract} \label{Abstract}
Optimal tracking in switched systems with controlled subsystem and Discrete-time (DT) dynamics is investigated. A feedback control policy is generated such that a) the system tracks the desired reference signal, and b) the optimal switching time instants are sought. For finding the optimal solution, approximate dynamic programming is used. Simulation results are provided to illustrate the effectiveness of the solution.
\end{abstract}
%%%%%%%%%%%%%%%%%%%%%%%%%%%%%%%%%%%%%%%%%%%%%%%%%%%%%%%%%%%%%%%%%%%%%%%%%%%%%%%%
%\begin{keyword}

%\textit{Index Terms-} optimal switching, fixed-mode sequence, controlled subsystems, approximate dynamic programming.
%\end{keyword}

\section{Introduction} \label{sec_intro}

In this study, switched systems are referred to as systems comprised of several subsystems/modes such that at each time instant only one subsystem is active. If the subsystems include continuous control signals, the subsystems are called \textit{controlled} subsystems. In case no control input exists in the subsystems, the subsystems are called \textit{autonomous} subsystems. Control of switched systems with controlled subsystem is a challenging problem as the controller needs to define the switching time and the continuous control in the subsystems. Also, if the sequence of active subsystems. i.e., mode sequence, is known a priori, the mode sequence is called a \textit{fixed mode sequence}. In a fixed mode sequence, the controller assigns the switching times and does not assign the active modes. 
%For example, in a switched system with 2 subsystems, one possible mode sequence is $\{mode\: 1, mode\: 2\}$. In this problem, the switching time is not known and it is the controller’s responsibility to define it. However, the controller does not need to decide on the mode sequence itself. 
On the other hand, if the mode sequence is free, then the controller needs to decide which mode should be activated at each time instant. Due to the discontinuous nature of the switched systems, control of these systems are challenging. Control of switched systems addresses many critical problems in automotive engineering \cite{Tohid_ABS_IET}, power engineering \cite{Ata_Heydari_DSCC, VamSwitch}, and many other engineering fields.  

%Optimal control is a branch of control that generates control signals by minimizing a cost function subjected to input/state constraints. 
From the mathematical point of view, solutions of the underlying Hamilton-Jacobi-Bellman (HJB) equation provide the necessary and sufficient condition for optimality in optimal control problems \cite{kirk2004optimal}. However,  solving the HJB equation explicitly is difficult and most cases impossible. 
%The solutions provided by variational calculus and minimum principle of Pontryagin are limited to systems with simple dynamics. 
Dynamic Programming (DP) can solve optimal control problems through discretizing the state and control domain and finding the optimal value functions backward in time. However, as the order of the system increases, rapid access to memory becomes prohibitive in DP which is known as \textit{curse of dimensionality}. To remedy the curse of dimensionality, one solution is to seek a near-optimal control solution instead of the exact optimal control which is provided by Approximate Dynamic Programming (ADP). In general, ADP methods use function approximators to approximate the value functions and they use iterative methods to tune the parameters of these function approximators.

ADP solutions for optimal control of switched systems with free mode sequence were investigated in \cite{HEYDARI20151620, HeydariSwiTra, SARDARMEHNI201810, Tohid_TNNLS, Wenji_TAC, Rienheart, VamSwitch}. As for the fixed mode sequence, a transformation was introduced in \cite{Xu_parametrization} to treat the switching times as parameters. This transformation was adapted in \cite{Heydari_Fixed_mode_seq} and the mode sequence was incorporated in the value functions to solve a regularization problem. 
%Also, \cite{Kamgarpour20121177} introduced a method for a certain class of switched systems which did not use the transformation in \cite{Xu_parametrization}. 
%For switched systems with autonomous subsystems, \cite{Seatzu1632302} introduced a DP-like solution. Also, direct search of the switching times instants with exhaustive search was introduced in \cite{SAKLY2009249}. 

To solve the optimal tracking problem in systems with conventional dynamics, some ADP solutions were introduced in \cite{Heydari_tracking, Kiumarsi_tracking_nonlinear}. In \cite{Heydari_tracking}, an ADP solution based on Single Network Adaptive Critic (SNAC) structure was introduced which approximated the optimal costates and used them to provide the optimal policy. Also, another ADP solution was developed in \cite{Kiumarsi_tracking_nonlinear} which approximated the value functions. For this method, a change of variables was performed in the system and a new state vector was introduced which included the tracking error dynamics and the dynamics of the reference signal. Furthermore, in \cite{Luo_tracking, 7001601} model free tracking methods were investigated. As for the switching dynamics, optimal tracking was studied in \cite{GAN20195193, SARDARMEHNI201810, HeydariSwiTra, Ata_Heydari_DSCC} in switched systems with autonomous subsystems and free mode sequence, and in \cite{ZHANG2019} in switched systems with controlled subsystems and free mode sequence.    

In this paper, optimal tracking in switched systems with controlled subsystems and fixed mode sequence is studied. The basic idea in this study is using the transformation introduced in \cite{Xu_parametrization} to parametrize the switching instants and then use the idea introduced in \cite{Heydari_Fixed_mode_seq} to include the mode sequence in the value functions and costates. Hence, the present study combines the results of \cite{Xu_parametrization, Heydari_Fixed_mode_seq, Heydari_tracking, Kiumarsi_tracking_nonlinear} to develop a solution for optimal tracking in switched systems with fixed mode sequence. To solve the optimal tracking problem, a single network is used to capture the optimal costates which are parametrized with respect to the switching instants. Once the optimal costates are known, some recommendations are given to find the optimal switching times from the costates at an initial condition.

The rest of the manuscript is organized as follows. In section \ref{sec_problem_formulation}, optimal control problem formulation and some assumptions are presented. In section \ref{sec_approach_1}, the proposed solution based on the SNAC is introduced. 
%The second approach is presented in section \ref{sec_approach_2}. 
Simulation results are discussed in section \ref{sec_simualtion}, and section \ref{sec_conclusion} concludes the paper.

\section{Problem Formulation} \label{sec_problem_formulation}

The dynamics of the switched systems can be shown as  
\begin{equation}
\begin{split}
\dot{x}(t)= \bar{f}_v\big(x(t)\big) + \bar{g}_v\big( x(t) \big) u(t), \: v\in \mathscr{V} = \{1,2,\dots, M\}, x(0) = x_0 \label{eq_dynamics} 
\end{split}
\end{equation}
where $x\in \mathbb{R}^n$ is the state vector, $u\in \mathbb{R}^m$ is the input, and $t$ denotes the time. The Lipschitz continuous functions $\bar{f}_v: \mathbb{R}^n \to \mathbb{R}^n$ and $\bar{g}_v: \mathbb{R}^n \to \mathbb{R}^{n\times m}$ denote the dynamics of the subsystems. The sub-index $v$ portrays the active mode which can be selected from the set of all available modes, $\mathscr{V}$, in the system. It is further assumed that $\bar{f}_v(0) = 0$, for all modes $ v\in \mathscr{V}$. The inclusion of continuous control, i.e., $u(.)$, in (\ref{eq_dynamics}) shows that the subsystems are controlled subsystems.  

Assuming the sequence of active modes is known, it is desired to find the continuous control policy $u(.)$, and the switching times, such that a performance index presented as 
\begin{equation}
\begin{split}
J&(x_0 )= \big(x(t_f)- r(t_f)\big)^T S \big( x(t_f) - r(t_f) \big) +  \int_{t_0}^{t_f} \frac{1}{2}\Big(\big(x(t)-r(t)\big)^T \bar{Q} \big(x(t)-r(t)\big)+ u(t)^T \bar{R} u(t) \Big) dt	\label{cost_func_tracking_contin}
\end{split}
\end{equation}
is minimized. In (\ref{cost_func_tracking_contin}), $t_0 $ is the initial time, and $t_f$ is the final time. $r \in \mathbb{R}^n$ is the reference signal where $\dot{r}(t) = \bar{f}_{r_v}(t)$ is the dynamics of the reference signal and the sub-index $r_v$ shows the active reference dynamics. $S \in \mathbb{R}^{n\times n}$ is a positive semi-definite matrix for penalizing the terminal cost, $\bar{Q}\in \mathbb{R}^{n\times n}$ is the state penalizing matrix which is assumed to be positive semi-definite, and $\bar{R} \in \mathbb{R}^{m \times m}$ is a positive definite control penalizing matrix.

Using Euler integration method, by choosing a small sample time $\delta t>0$, one can discretize the dynamics (\ref{eq_dynamics}) as
\begin{equation}
x_{k+1} = f_v(x_k) + g_v(x_k) u_k \label{dynamics_discrete}
\end{equation}
where the non-negative integer $k$ is the discrete time index. For notational simplicity, the discrete time index is shown as a sub-index, i.e., $x_k \equiv x(k) $. Also, $f_v(x_k) = x_k + \bar{f}_v(x_k) \delta t$, and $g_v(x_k) = \bar{g}_v(x_k) \delta t$. With a similar procedure, one can discretize the cost function (\ref{cost_func_tracking_contin}) as 
\begin{equation} 
\begin{split}
J&(x_0)= (x_N - r_N)^T S ( x_N - r_N )  +  \sum_{k = 0 }^{N} \frac{1}{2}\Big(\big(x_k-r_k\big)^T Q \big(x_k-r_k\big)+ u_k^T R u_k\Big) \label{cost_func_tracking_discrete}
\end{split}
\end{equation}
In (\ref{cost_func_tracking_discrete}), $N = \frac{t_f - t_0}{\delta t}$, $Q = \bar{Q} \delta t$, and $R = \bar{R} \delta t$. Based on (\ref{cost_func_tracking_discrete}), one can define the cost-to-go as the cost of going from discrete time index $k$ to $N$ as 
\begin{equation} 
\begin{split}
J&(x_k)= \big(x_N- r_N\big)^T S \big( x_N - r_N \big) +  \sum_{\bar{k} = k }^{N} \frac{1}{2}\Big( (x_{\bar{k}}-r_{\bar{k}})^T Q (x_{\bar{k}}-r_{\bar{k}})+ u_{\bar{k}}^T R u_{\bar{k}}\Big) \label{cost_to_go}
\end{split}
\end{equation}
Before going forward, the following definition and assumption are needed. 
\begin{Def} \label{admissibility}
A control policy is called admissible if it stabilizes the system presented in (\ref{dynamics_discrete}) in a selected compact region of interest $\Omega \subset \mathbb{R}^n$, which includes the origin. Also, $\forall x_0 \in \Omega$, $J(x_0)$ should be finite if the state is propagated along that policy.
\end{Def}

\begin{Assumption} \label{ass_existence_admin_policy}
Given the mode sequence, there is at least one admissible policy for the system. 
\end{Assumption}

Considering Assumption \ref{ass_existence_admin_policy}, one can define the value function as 
\begin{equation} 
\begin{split}
V&\big(x_k, k\big)\equiv V_k(x_k)= \min_{u(.)}\biggr((x_N- r_N)^T S (x_N - r_N) + \frac{1}{2}\sum_{\bar{k} = k }^{N} \Big( (x_{\bar{k}}-r_{\bar{k}})^T Q (x_{\bar{k}}-r_{\bar{k}})+ u_{\bar{k}}^T R u_{\bar{k}}\Big) \biggr) \label{cost_to_go_1}
\end{split}
\end{equation}
Considering time step $k$ to $k+1$, one has 
\begin{equation} 
\begin{split}
&V_k(x_k)= \min_{u(.)}\Big((x_N- r_N\big)^T S \big( x_N - r_N \big) + \frac{1}{2}(x_k-r_k)^T Q (x_k-r_k)+ \frac{1}{2}u_k^T R u_k + \frac{1}{2}\sum_{\bar{k} = k+1 }^{N} (x_{\bar{k}}-r_{\bar{k}})^T Q (x_{\bar{k}}-r_{\bar{k}})+ u_{\bar{k}}^T R u_{\bar{k}}\Big) \label{cost_to_go_2}
\end{split}
\end{equation}
After some algebraic manipulations, one can rewrite (\ref{cost_to_go_2}) as
\begin{equation} 
\begin{split}
V_k(x_k)= \min_{u(.)}\Big(&\frac{1}{2}(x_k-r_k)^T Q (x_k-r_k)+ \frac{1}{2}u_k^T R u_k + V_{k+1}(x_{k+1})\Big) \label{cost_to_go_3}
\end{split}
\end{equation} 
Equation (\ref{cost_to_go_3}) simply means minimum cost of going from time $k$ to $N$ equals to cost of going from time $k$ to $k+1$ plus minimum cost of going from time $k+1$ to $N$. This is in fact the Bellman equation of optimality \cite{kirk2004optimal}. Based on (\ref{cost_to_go_3}), one can define the optimal policy as 
\begin{equation} 
\begin{split}
u_k(x_k)= \argmin_{u(.)}\Big(&\frac{1}{2}(x_k-r_k)^T Q (x_k-r_k)+ \frac{1}{2}u_k^T R u_k + V_{k+1}(x_{k+1})\Big) \label{optimal_policy_3}
\end{split}
\end{equation} 
%
%%%%%%%%%%%%%%%%%%%%%%%%%%%%%%%%%%%%%%%%%%%%%%%%%%%%%%%%%%%%%%%%%%%%%%%%%%%%%%%%%%%
%%%%%%%%%%%%%%%%%%%%%%%%%%%%%%%%%%%%%%%%%%%%%%%%%%%%%%%%%%%%%%%%%%%%%%%%%%%%%%%%%%%
%%%%%%%%%%%%%%%%%%%%%%%%%%%%%%%%%%%%%%%%%%%%%%%%%%%%%%%%%%%%%%%%%%%%%%%%%%%%%%%%%%%
%%%%%%%%%%%%%%%%%%%%%%%%%%%%%%%%%%%%%%%%%%%%%%%%%%%%%%%%%%%%%%%%%%%%%%%%%%%%%%%%%%%
%%%%%%%%%%%%%%%%%%%%%%%%%%%%%%%%%%%%%%%%%%%%%%%%%%%%%%%%%%%%%%%%%%%%%%%%%%%%%%%%%%%
%%%%%%%%%%%%%%%%%%%%%%%%%%%%%%%%%%%%%%%%%%%%%%%%%%%%%%%%%%%%%%%%%%%%%%%%%%%%%%%%%%%
\subsection{Including the Mode Sequence}\label{sub_sec_including_mode_seq}

In \cite{Xu_parametrization}, a transformation is introduced to include the switching times as parameters in the optimal control formulation. For ease of presentation, a case with two subsystems and only one switching is considered. Let the switching happen at $t = t_1\in (t_0,t_f)$. Also, let the mode sequence to be $\{mode$ $1,$ $mode$ $2\}$\footnote{When $t< t_1$ mode 1 is active and when $t\ge t_1$ mode 2 is active. }. To make the switching time instant an independent parameter, let \cite{Xu_parametrization} 
\begin{equation}
t =
\begin{cases*}
t_0 + (t_1 - t_0) \hat{t} & if $0 \leq \hat{t}< 1$ \\
t_1 + (t_f - t_1) (\hat{t} - 1) & if $1\leq \hat{t} \leq 2$    
\end{cases*}
\label{eq_transformed_time}
\end{equation}
From the transformation introduced in (\ref{eq_transformed_time}), one notices that $t \in [t_0, t_f]$ and $\hat{t} \in [0,2]$. The merit of the transformation is that the switching time $t_1$ can be any point in $t\in [t_0, t_f]$. However, in the transformed time, i.e., $\hat{t} \in [0,2]$, switching only happens at $\hat{t} = 1$. For $\hat{t} \ge 1$, mode 2 is active and for $\hat{t} < 1$, mode 1 is active. Based on the introduced transformation in (\ref{eq_transformed_time}), one has
\begin{equation} 
x'(\hat{t}) = \frac{dx}{d \hat{t} } =  \frac{dx}{dt} \frac{dt}{d\hat{t}} \label{diff_trans}
\end{equation}
Since the mode sequence is known, (\ref{diff_trans}) becomes
\begin{equation}
x'(\hat{t})  =
\begin{cases*}
\Big(\bar{f}_1\big(x(\hat{t})\big) + \bar{g}_1\big(x(\hat{t})\big)u(\hat{t}) \Big) (t_1 - t_0) & if $0 \leq \hat{t}< 1$ \\
\Big(\bar{f}_2\big(x(\hat{t})\big) + \bar{g}_2\big(x(\hat{t})\big)u(\hat{t}) \Big) (t_f - t_1) & if $1\leq \hat{t} \leq 2$    
\end{cases*}
\label{eq_transformed_dynamics}
\end{equation}
Also, the cost function in (\ref{cost_func_tracking_contin}) can be written as\footnote{Since the mode sequence is known, one can consider the integral from $t_0$ to $t_1$ with the first mode, and from $t_1 $ to $t_f$ with the second mode. }
\begin{equation} 
\begin{split}
J(x_0)  =& \big(x(2) - r(2)\big)^T S \big(x(2) - r(2)\big) + \int_{0}^{1} \frac{1}{2}\Big( \big(x(\hat{t}) - r(\hat{t})\big)^T \bar{Q} (t_1 - t_0) \big(x(\hat{t})- r(\hat{t})\big) + u(\hat{t})^T \bar{R}(t_1 - t_0) u(\hat{t}) \Big)d\hat{t}\\
& + \int_{1}^{2} \frac{1}{2}\Big( \big( x(\hat{t})- r(\hat{t})\big)^T \bar{Q} (t_f - t_1) \big(x(\hat{t})- r(\hat{t})\big) + u(\hat{t})^T \bar{R}(t_f - t_1) u(\hat{t}) \Big)d\hat{t}
\label{cost_switching_transformed}
\end{split}
\end{equation}
An important observation in (\ref{cost_switching_transformed}) is that the transformed cost function is not only a function of $x_{0}$, but also it is a function of the switching time, i.e., $t_1$. Hence, $J(x_0) = J(t_1, x_0)$. With a similar procedure used before, by choosing a small sampling time $\delta \hat{t}$ one can discretize (\ref{eq_transformed_dynamics}) and (\ref{cost_switching_transformed}) as
\begin{equation}
x_{\hat{k}+1} =
\begin{cases*}
f_1\big(x_{\hat{k}} \big)  + g_1\big(x_{\hat{k}}\big) u_{\hat{k}}  & if $0 \leq \hat{k}< \frac{N'}{2}$ \\
f_2\big(x_{\hat{k}} \big)  + g_2\big(x_{\hat{k}}\big) u_{\hat{k}}  & if $\frac{N'}{2} \leq \hat{k}\leq N'$     
\end{cases*}
\label{eq_transformed_dynamics_for_lambda_DT_1}
\end{equation}
where $f_1(x_{\hat{k}} ) =  x_{\hat{k}} +  \bar{f}_1(x_{\hat{k}}) (t_1 - t_0) \delta \hat{t}$, $g_1(x_{\hat{k}} ) =  \bar{g}_1(x_{\hat{k}} ) (t_1 - t_0) \delta \hat{t}$, $f_2(x_{\hat{k}} ) =  x_{\hat{k}} +   \bar{f}_2(x_{\hat{k}}) (t_f - t_1) \delta \hat{t}$, and $g_2(x_{\hat{k}} ) = \bar{g}_2(x_{\hat{k}}) (t_f - t_1) \delta \hat{t}$. In (\ref{eq_transformed_dynamics_for_lambda_DT_1}), $\hat{k} \in [0 ,N']$ is the discrete time index where $N' = \frac{number\hspace{3 pt} of \hspace{3 pt}switching + 1}{\delta \hat{t}}$ \cite{Heydari_Fixed_mode_seq}. 
With the transformed dynamics and the cost function, with similar procedure used in the previous section one can define cost-to-go as

\begin{equation} 
V_{\hat{k}}(t_1, x_{\hat{k}}) =  
\begin{cases*}
\mathbf{Q}_1 + \mathbf{R}_1 + V_{\hat{k}+1}(t_1, x_{\hat{k}+1} )& if $0 \leq \hat{k}< \frac{N'}{2}$ \\
\mathbf{Q}_2 + \mathbf{R}_2 + V_{\hat{k}+1}(t_1, x_{\hat{k}+1} )& if $\frac{N'}{2} \leq \hat{k}\leq N'$ 
\label{cost_to_go_trans_swit}
\end{cases*}
\end{equation}
where
\begin{equation}
\begin{split}
\mathbf{Q}_1 &= \frac{1}{2}( x_{\hat{k}}- r_{\hat{k}} )^T \bar{Q} (t_1 - t_0) \delta \hat{t} ( x_{\hat{k}}- r_{\hat{k}})\\
\mathbf{R}_1 &= \frac{1}{2}u_{\hat{k}}^T \bar{R}(t_1 - t_0)\delta \hat{t} u_{\hat{k}} \\
\mathbf{Q}_2 &= \frac{1}{2}( x_{\hat{k}}- r_{\hat{k}} )^T \bar{Q} (t_f - t_1) \delta\hat{t} ( x_{\hat{k}}- r_{\hat{k}})\\
\mathbf{R}_2 &= \frac{1}{2}u_{\hat{k}}^T \bar{R}(t_f - t_1)\delta \hat{t} u_{\hat{k}}
\end{split}
\end{equation}
As one can see, the cost-to-go in (\ref{cost_to_go_trans_swit}) is a function of current time $\hat{k}$, current state $x_{\hat{k}}$, and the switching time $t_1$. Similarly, one can define the costate as\footnote{By definition, costate is the gradient of value function, i.e., $\lambda(x) = \frac{\partial V(x)}{\partial x}$. Taking the gradient of (\ref{cost_to_go_trans_swit}) leads (\ref{costate_trans_swit}).}
\begin{equation} 
\lambda_{\hat{k}}(t_1, x_{\hat{k}}) =  
\begin{cases*}
\bar{\mathscr{Q}}_1 + \frac{\partial x_{\hat{k}+1}}{\partial x_{\hat{k}}}\lambda_{\hat{k} + 1}(t_1, x_{\hat{k}+1})& if $0 \leq \hat{k}< \frac{N'}{2}$ \\
\bar{\mathscr{Q}}_2 + \frac{\partial x_{\hat{k}+1}}{\partial x_{\hat{k}}}\lambda_{\hat{k}+1}(t_1, x_{\hat{k}+1})& if $\frac{N'}{2} \leq \hat{k}\leq N'$ 
\label{costate_trans_swit}
\end{cases*}
\end{equation}
where 
\begin{equation}
\begin{split}
\bar{\mathscr{Q}}_1 &= \bar{Q} (t_1 - t_0) \delta \hat{t} \big( x_{\hat{k}}- r_{\hat{k}}\big)\\
\bar{\mathscr{Q}}_2 &= \bar{Q} (t_f - t_1) \delta \hat{t} \big( x_{\hat{k}}- r_{\hat{k}}\big)
\end{split}
\end{equation}

\section{Single Network Adaptive Critic (SNAC)} \label{sec_approach_1}

The application of SNAC was introduced for tracking in systems with conventional dynamics \cite{Heydari_tracking}. This idea is adapted in this section to perform tracking in a switched system. To introduce the concept, consider the costate as in (\ref{costate_trans_swit}). 
%Equation (\ref{costate_trans_swit}) can be iterated one step to get the recursive relationship for the costate. 
The idea here is training neural networks to approximate $\lambda_{\hat{k}+1}(t_1, x_{\hat{k}+1})$ from $(t_1,x_{\hat{k}})$. Based on Weierstrass Approximation Theorem \cite{Rudin}, linear-in-parameter neural networks with polynomial basis functions can uniformly \cite{HornikFeedforward} approximate continuous functions to a desired degree of precision in a compact set. In order to use Weierstrass Approximation Theorem, the following assumption is required. 
\begin{Assumption} \label{ass_contin_diffable}
The value functions are continuously differentiable.% with respect to the states, i.e., $x_{\hat{k}}$. 
\end{Assumption}
Through Assumption \ref{ass_contin_diffable}, one can use linear-in-parameter neural networks to approximate the value functions, and the costates. Consider the exact costate at discrete time index $\hat{k}$ as 
\begin{equation}
\lambda_{\hat{k}+1}(t_1, x_{\hat{k}+1}) = {W_{\hat{k}}^*}^T \phi(t_1, x_{\hat{k}}) + \varepsilon_{\hat{k}}^*(t_1, x_{\hat{k}}) \label{exact_lambda_function}
\end{equation} 
where $W_k^* \in \mathbb{R}^{m_{\lambda} \times n}$ is a weight vector and $\phi: \mathbb{R} \times \mathbb{R}^n \to \mathbb{R}^{m_{\lambda}}$ is a vector of linearly independent polynomial basis functions (neurons). The number of neurons is denoted by positive integer $m_{\lambda}$. In (\ref{exact_lambda_function}), the dependence of the parameters/functions to discrete time index is shown with a sub-index $\hat{k}$. Hence the approximate costates can be calculated as 
\begin{equation} 
\widehat{\lambda}_{\hat{k}+1}(t_1, x_{\hat{k}+1}) = \widehat{W}_{\hat{k}}^T \phi(t_1, x_{\hat{k}})
\label{approximate_costate}
\end{equation}
where $\widehat{W}_{\hat{k}} \in \mathbb{R}^{m_{\lambda} \times n}$ is a tunable weight vector. The $\widehat{W}_{\hat{k}}$ is tuned through the training process. Once the costates are known, one finds the optimal policy as 
\begin{equation} 
\widehat{u}_{\hat{k}}(t_1, x_{\hat{k}}) =  
\begin{cases*}
-R_1^{-1}g_1^T\big(x_{\hat{k}}\big) \widehat{\lambda}_{\hat{k}+1}(t_1, x_{\hat{k}+1})& if $0 \leq \hat{k}< \frac{N'}{2}$ \\
-R_2^{-1}g_2^T\big(x_{\hat{k}}\big) \widehat{\lambda}_{\hat{k}+1}(t_1, x_{\hat{k}+1})& if $\frac{N'}{2} \leq \hat{k}\leq N'$ 
\label{Policy_trans_swit}
\end{cases*}
\end{equation}
In (\ref{Policy_trans_swit}), $R_1= R \delta \hat{t}(t_1 - t_0) $ and $R_2 = R \delta \hat{t}(t_f - t_1)$. For training, one can go backward in time and find the costates and save them for online control. This process is summarized in Algorithm \ref{alg_training_costate_switching}.

\begin{algorithm}
\caption{\textbf{:} \textit{Finding the Costates (Approach 1)}}
\label{alg_training_costate_switching}
\begin{algorithmic}[1]
\item [\textbf{step 1:}] Initialize the neural network weights, $\widehat{W}^0_{\hat{k}}, \hat{k} \in \{ 1, \dots, N-1 \}$. Also select a small positive number $\gamma$ as a convergence tolerance. Select $\eta$ random training samples $x^{[l]}\in \Omega$ where $l \in \{ 1, 2, ..., \eta\} $ and $t_1^{[l]}\in [t_0, t_f]$ where $l \in \{ 1, 2, ..., \eta\}$. 
\item [\textbf{step 2:}] Repeat the outer loop for $\hat{k} = N-1$ to $\hat{k} = 1$:
\item [\textbf{step 3:}] Set $i =0$ and repeat the following inner loop: 
\item [\textbf{step 3-1:}] Select $\eta$ random training samples $\{(t_1^{[l]}, x^{[l]})\in (t_0, t_f) \times \Omega$ where $l \in \{ 1, 2, ..., \eta\}$. Substitute all the training samples in $\phi(.,.)$ and find a $\lambda_{\hat{k}+1}(t_1, x_{\hat{k}})$. With $\lambda_{\hat{k}+1}(t_1, x_{\hat{k}})$ find $u_{\hat{k}}$ and propagate the states along it to find $x_{\hat{k}+1}$. Also, find $r_{\hat{k}+1}$. 
\item [\textbf{step 3-2:}] If $\hat{k} = N-1$ set the target as $ \widehat{\lambda}_{\hat{k}+1}(t_1, x_{\hat{k}+1}) = S (x_{N'} - r_{N'})$ and train $\widehat{W}^{i+1}_{\hat{k}}$ such that $\widehat{W}^{{i+1}^T}_{\hat{k}} \phi(t_1, x_{\hat{k}}) = \widehat{\lambda}_{\hat{k}+1}(t_1, x_{\hat{k}+1}) $ with least squares on the entire set of training samples. Otherwise, set the target as right-hand side of (\ref{costate_trans_swit}) and use the least squares to find $\widehat{W}^{{i+1}^T}_{\hat{k}}$. 
\item [\textbf{step 3-3:}] If $\| \widehat{W}^{i+1}_{\hat{k}} - \widehat{W}^{i}_{\hat{k}}\| \leq \gamma$, go to step 4. Otherwise, set $i=i+1$ and go back to step 3-1. 
\item [\textbf{step 4:}] If $\hat{k} =1$, stop the training. Otherwise, store $\widehat{W}^{i+1}_{\hat{k}}$, set $\hat{k} = \hat{k}-1$ and go to step 2.  
\end{algorithmic}
\end{algorithm}

\begin{Rem}\label{covergence_inner_loop_approach_1}\normalfont
The convergence of the inner loop in step 3 of Algorithm \ref{alg_training_costate_switching} was studied in theorem 1 of \cite{Heydari_tracking} for systems with conventional dynamics. The proof can be modified to include tracking in switched systems. 
\end{Rem}

\begin{Rem}\label{fining_swtiching_time_from_costates}
\normalfont

Once the training is concluded, one needs to find the optimal switching times from the costates for a selected initial condition $x_0 \in \Omega$. Three methods are suggested below to find the optimal switching times from the optimal costates.  

\begin{itemize}
\item \textbf{-- Method 1:} propagating the states analytically along the optimal policy by treating switching time as a parameter and finding the optimal cost-to-go from the cost function. Once done, one can use constrained minimization methods to find switching times. 
\item \textbf{-- Method 2:} integrating the costate analytically to find the value function. %This method is suitable for systems with low order dynamics. 
Similar to finding the velocity field from potential flow in fluid mechanics, one can integrate the costates analytically to find the value functions. The convenient feature of this method is that the analytical solutions provide the optimal value function $\forall x_0 \in \Omega$. In order word, one does not need to integrate again when the initial condition is changed. This is unlike Method 1 that propagation should be done for each initial condition separately.  However, as the order of system increases, this method becomes very complicated. In other words, this method is only suitable for systems with low order dynamics.  
\item \textbf{-- Method 3:} propagating the states along all possible switching times and find the optimal cost to go for all possible switching time. Once done, choose the switching times which lead to the minimum value function. Method 3 is similar to the forward dynamic programming method and when the number of switching increases, performing this method might become very time-consuming. 
\end{itemize}

\end{Rem}

\section{Numerical Simulation} \label{sec_simualtion}

In this section some simulation results are provided to evaluate the effectiveness of the solutions discussed in this paper. Consider a system with two modes. For the first mode, a benchmark system, Van der Pol oscillator was selected. The dynamics of this mode can be shown as 
\begin{equation}
\begin{split} 
\dot{x}_1(t) &= x_2(t)\\
\dot{x}_2(t) &= \big(1 - x^2_1(t) \big)  x_2(t) - x_1(t) + u(t)
\label{vad_der_pal}
\end{split}
\end{equation} 
For the second mode, a linear subsystem was selected as
\begin{equation}
\begin{split} 
\dot{x}_1(t) &= x_2(t)\\
\dot{x}_2(t) &= 2x_1 - x_2 + u(t)
\label{sub_sys_2}
\end{split}
\end{equation} 
The reference signal is chosen as 
\begin{equation}
\begin{split} 
\dot{r}_1(t) &= \sin(\pi t)\\
\dot{r}_2(t) &= \pi \cos(\pi t)
\label{reference_sig}
\end{split}
\end{equation}  

\begin{figure}[htbp]
\begin{center}
\includegraphics[scale = 0.9]{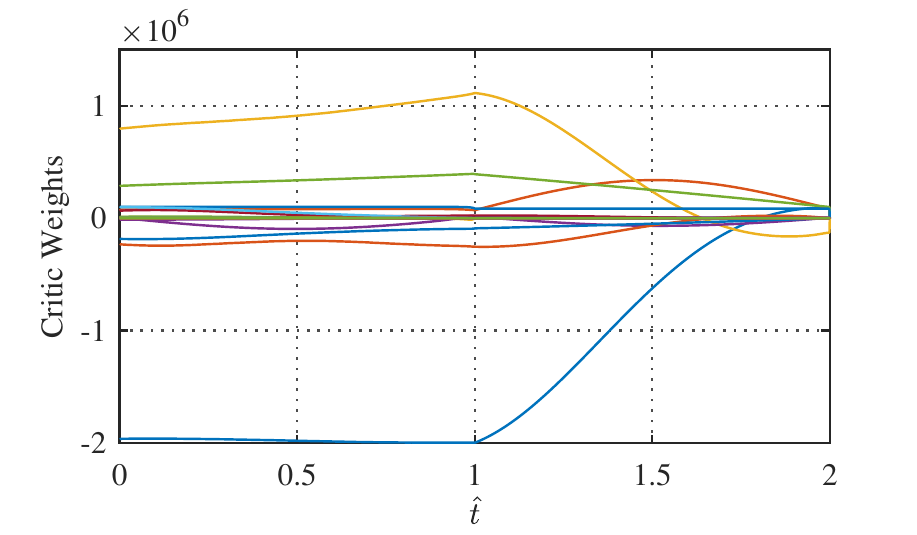}
\end{center}
\vspace{5pt}	
\caption{The history of the  weights of the neural networks to approximate costates. The jump at $\hat{t} = 1$ shows the switching at this time.} 
%\captionsetup{justification=centering}
\label{critic_weights_transformed}
\vspace{-10pt}
\end{figure}

For the cost function, $S = \diag(10^5 , 10^5)$, $\bar{Q} = \diag(10^5, 10^7)$ were selected, where $\diag(a,b)$ denotes a diagonal matrix with $a$ and $b$ on the main diagonal and $0$ elsewhere. Also, $\bar{R} = 1/\delta \hat{t} $ was selected as the control penalizing term. The basis functions were selected as all possible combination of $t_1$, $x_1$, $x_2$ up to the power 3 without repetition. 
%
%Also, the following neurons were included as $x_1^b x_1^c$ where $b+c = 4$. 
For training, 1000 random samples were generated in $\Omega = \{ (t_1, x_1, x_2) | t_1\in (0,3), x_1\in [-4,4],  x_2 \in [-4,4] \}$. For descritization, sampling time was selected as $\delta \hat{t} = 0.001$ ($sec$). Using approach 1, The training process concluded in $22.6177$ seconds using an office desktop with 16 GB of RAM and Intel(R) Core(TM) i7-3770 CPU @ 3.4 GHz. The history of the weights of the neural networks to approximate the costates is shown in Fig. \ref{critic_weights_transformed}. As one can see from Fig \ref{critic_weights_transformed}, the history of the weights shows a jump at $\hat{t} =1$ which is the switching time.  

Once the training concluded, the trained costates were used to find the optimal switching time. For this example, Method 2 in Remark \ref{fining_swtiching_time_from_costates}, i.e., integration of the costate analytically, was used to find the value function and then minimized the value function with nonlinear programming methods. Through integration, one can find the analytical value function as $V(t_1,x_1, x_2) =67377 t_1^3 x_1 + 568.7 t_1^3 x_2 + 49288 t_1^2 x_1^2 + 849 t_1^2x_1x_2 - 2.322\times 10^5t_1^2x_1 + 57.67t_1^2x_2^2 - 2077t_1^2x_2 + 4210t_1x_1^3 + 162.2t_1x_1^2x_2 - 92177t_1x_1^2 + 1.432t_1x_1x_2^2 - 1691t_1x_1x_2 + 7.968\times 10^5t_1x_1 + 0.3332t_1x_2^3 - 2411t_1x_2^2 + 3337t_1x_2 + 2413x_1^4 + 97.95x_1^3x_2 - 4013x_1^3 - 434.8x_1^2x_2^2 - 182.2x_1^2x_2 + 1.44\times 10^5x_1^2 - 2.424x_1x_2^3 + 2.186x_1x_2^2 + 1789x_1x_2 - 1.961\times 10^6x_1 - 0.1953x_2^4 - 0.08602x_2^3 + 50288x_2^2 + 77666x_2
$. Substituting the initial conditions for $x_1 = 1$ and $x_2 = -0.5$, one can find the value function as a function of the switching time as $V(t_1) = 67088t_1^3 - 1.823 \times 10^5 t_1^2 + 7.073 \times 10^5 t_1 - 1.846 \times 10^6$. Using nonlinear programming, the best switching time is sought as $t_1 = 2.654$ ($sec$). 

\begin{figure}[htbp]
\begin{center}
\includegraphics[scale = 0.9]{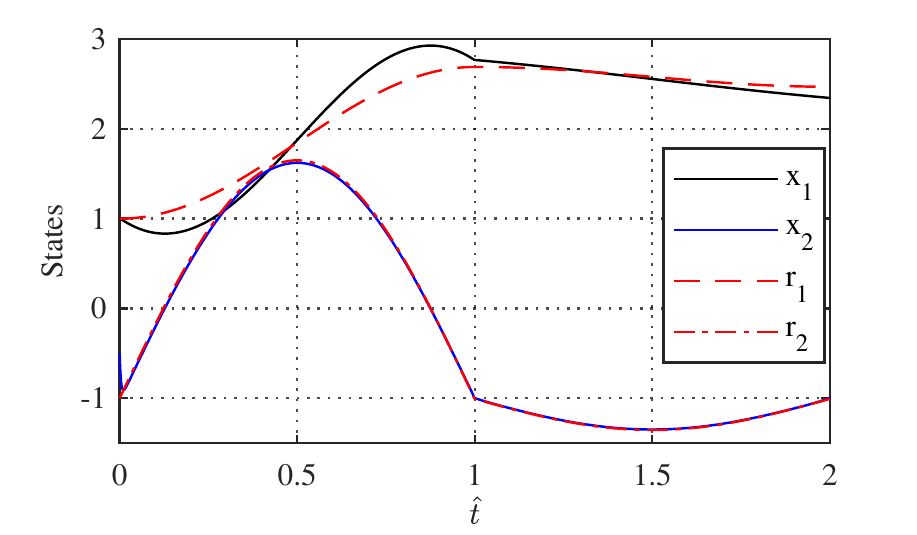}
\end{center}
\vspace{-10pt}	
\caption{The history of the states in the transformed time. }
%\captionsetup{justification=centering}
\label{states_transformed}
\vspace{-8pt}
\end{figure}

The history of the states in the transformed dynamics is shown in Fig. \ref{states_transformed}. As one can see, the controller had a very good performance in forcing $x_2(.)$ to track $r_2(.)$ as in the cost function it was emphasized.  

\section{Conclusion} \label{sec_conclusion}
An approximate solution for optimal control of switched systems with fixed mode sequence and controlled subsystems was presented. The method includes two levels of control. In the upper level, optimal switching times were sought. In the lower level, continuous control for each mode was generated in a feedback form. To find the continuous control, a single network adaptive critic was used to find the optimal costates while treating the switching times as parameters. Simulation results confirmed the effectiveness of the solution.

\bibliography{Reference_Report_2}
\bibliographystyle{IEEEtran}
\end{document}